\documentclass[unnumsec,webpdf,contemporary,large]{oup-authoring-template}%
\usepackage{graphicx}%
\usepackage{float} %
\usepackage{subfigure}%
\usepackage{caption}
\usepackage{algorithm}  
\usepackage{algpseudocode}  
\usepackage{amsmath}  
\renewcommand{\algorithmicrequire}{\textbf{Input:}} 
 
\usepackage{arydshln} 
\usepackage{longtable}
\usepackage{makecell}
\usepackage[section]{placeins}
\usepackage{amsmath}
\usepackage{soul,color}

\graphicspath{{Fig/}}

\theoremstyle{thmstyleone}%
\theoremstyle{thmstyletwo}%
\theoremstyle{thmstylethree}%

\begin{document}

\journaltitle{}
\DOI{DOI HERE}
\copyrightyear{2023}
\pubyear{2023}
\access{Advance Access Publication Date: Day Month Year}
\appnotes{Paper}

\firstpage{1}


\title[BOSS]{BOSS - Biomarker Optimal Segmentation System}

\author[1]{Liuyi Lan}
\author[2]{Xuanjin Cheng}
\author[3,$\ast$]{Li Xing \ORCID{0000-0002-4186-7909}}
\author[1,$\ast$]{Xuekui Zhang \ORCID{0000-0003-4728-2343}}

\authormark{Liuyi et al.}

\address[1]{\orgdiv{Department of Mathematics and Statistics}, \orgname{University of Victoria}, \orgaddress{\state{Victoria, BC}, \country{Canada}}}
\address[2]{\orgdiv{Department of Medical Genetics}, \orgname{The University of British Columbia}, \orgaddress{\state{Vancouver, BC}, \country{Canada}}}
\address[3]{\orgdiv{Department of Mathematics and Statistics}, \orgname{University of Saskatchewan}, \orgaddress{\state{Saskatoon, SK}, \country{Canada}}}

\corresp[$\ast$]{Xuekui Zhang. \href{ubcxzhang@gmail.com}{ubcxzhang@gmail.com} and Li Xing\href{lix491@mail.usask.ca}{lix491@mail.usask.ca} }

\received{Date}{0}{Year}
\revised{Date}{0}{Year}
\accepted{Date}{0}{Year}


\abstract{
\textbf{Motivation:} Precision medicine is a major trend in the future of medicine. It aims to provide tailored medical treatment and prevention strategies based on an individual's unique characteristics and needs. Biomarker is the primary source of patients' unique features used in precision medicine. We often need to investigate many cutoff values of a continuous biomarker to find the optimal one and test if it can help segment patients into two groups with significantly different clinical outcomes. This requires multiple testing adjustment on tests conducted on overlapped data. 
The permutation-based approach is often a preferred solution, since it does not suffer the limitations of state-of-art theoretical methods. However, permutation is computationally expensive and limits its application scenarios, such as web applications requiring a fast response or the analysis of genomic study requiring to repeat analysis many times on tens of thousands of genes.\\
\textbf{Results:} We proposed a novel method BOSS, Biomarker Optimal Segmentation System, to solve this problem. In simulation studies, we found BOSS’s statistical power and type I error control are both non-inferior to the permutation approach, and it is hundreds of times faster than permutation. To illustrate our method, we applied BOSS to real data and revealed potentially converging biomarkers that have referential importance in exploring synergy and target-matched therapies in lung adenocarcinoma. \\
\textbf{Availability:} An R package \textit{boss} is being developed and will be available on \url{https://cran.r-project.org/}. 
}


\maketitle

\section{Introduction}
Precision medicine, the medical approach of employing patients' specific information to tailor medical treatments to satisfy their needs, has become increasingly popular recently. It is partially due to the contribution of technological advances, enabling individuals' genomic data collection at a fast speed with reduced cost (add a reference). And these genomic data serve as the essential and major source for biomarkers in precision medicine \citep{Ashina, Vargas}. One of the most common practices of implementing precision medicine in health care and management is to use biomarkers to stratify patients into subgroups according to their associated risks of a disease or their likelihood of positive responses to treatments or interventions. Different treatments are applied to the corresponding subgroups based on such stratification to alleviate risks or improve positive responses. In practice, we often seek a cutoff value of a continuous biomarker to segment patients into two subgroups labelled as high-risk vs low-risk groups or responder vs non-responder and assign different treatments to the two groups. To achieve this goal, we must investigate multiple candidate cutoff values and answer two closely related questions, (1) which cutoff is the optimal one, and (2) whether the outcome variables differ significantly between these two groups segmented using the biomarker. And the answers can be revealed based on testing a series of candidate cutoffs to check their significance and compare their importance based on the effects carried on the test statistics. In this situation, we must apply multiple testing adjustments to correct the inflated false positive probabilities due to conducting multiple testings.

The popular multiple testing adjustment methods control the false discovery rate (FDR) or the familywise error rate (FWER). FDR is defined as the expected proportion of false rejections among the total rejections, considering the seriousness of the loss incurred by erroneous rejections to be inversely related to the number of hypotheses rejected \citep{Benjamini}. FWER is defined as the probability of making any false rejections. When a small number of false positives do not affect the model's overall effectiveness, controlling FDR is often preferred since it is less stringent and leads to more statistical power than FWER. So FDR is much more commonly used when multiple testing adjustments are needed. However, the FDR adjustment method requires a large number of hypothesis testing to estimate the FDR precisely. In the problem of finding the optimal cutoff, the more candidate cutoff values investigated, the heavier multiple testing penalties are introduced, leading to more severe lack-of-power issues. Thus, we often cannot investigate enough candidate cutoff values to estimate FDR reliably. Hence, we choose to control FWER in this problem.

Bonferroni correction \citep{ Bonferroni} is the most popular FWER control method. To control FWER below a threshold $\alpha$, Bonferroni correction uses $\alpha/n$ as the significance level for all tests, where $n$ is the number of hypotheses to be tested. As it assumes all hypotheses are independent, Bonferroni correction is often over-conservative and reduces statistical power. For example, when we repeatedly conduct the `same' test on the `same' dataset twice, we will get the same p-value twice, and we should not apply any multiple testing adjustments to this p-value even though two tests are carried out. The rule can be further generalized. The penalty applied to multiple testing should be relied on the correlations between these tests. Less penalty should be considered under high correlations among the test statistics. Different candidate cutoff converts the continuous biomarker into different binary variables in the segmentation problem. Any two derived binary variables overlap, and patients whose biomarker levels are above or under both cutoffs have exactly the same 0/1 values in the derived variables. Consequently, the corresponding test statistics obtained from these derived variables must be correlated. The closer the values of the two cutoffs, the higher the correlation between their test statistics. Therefore, Bonferroni correction is unsuitable for our problem due to its independence assumption. Worsley \citep{Worsley} proposed an improved Bonferroni, which orders the datasets and explicitly models correlations between test statistics obtained from every two `adjacent' datasets. However, this method assumes that test statistics obtained from non-adjacent datasets are independent, which is still unrealistic in our problem.

Meanwhile, advanced methods are developed using the maximally selected statistic and its asymptotic distribution to select the optimal cutoffs and adjust for multiple tests. For binary outcomes, Miller and Siegmund \cite{MSC} proposed to use a maximally selected chi-squared statistic to seek optimal cutoff and Boulesteix and Strobl\cite{MSCS} proved that using standard chi-square distribution can render more significance in coefficient estimation and proposed to obtain true critical values by deriving the asymptotic Gaussian distribution of maximally chi-square statistic. For continuous and time-to-event outcomes, Lausen and Schumacher\cite{Lausen1992} suggested the standardized maximally selected rank statistic. However, these methods are only applicable under a large number of candidate cutoffs due to the requirement of asymptotic properties. As mentioned before, there cannot be many cutoffs in our case. To adequately address these issues with fewer candidate cutoffs, Hothorn and Zeileis\cite{STA} proposed the permutation method, which randomly shuffles part of the data to estimate the empirical distribution of the maximally selected statistic test under the null hypothesis instead of employing their asymptotic distribution. Therefore, their method is still valid for the nonlarge number of candidate cutoffs. Hilsenbeck and Clark\cite{HILSENBECK} compared multiple testing adjustments based on the improved Bonferroni and the use of asymptotic distribution and empirical permutation of the maximally selected statistic via simulation studies and showed that the rejection rates from the permutation method are most close to those of the independent validation sample. Hilsenbeck et al. \cite{Susan} also noted that permutation and resample can better identify prognostic factors than other methods regarding accuracy and efficiency. Because of the advantages of the permutation methods, Cheng et al. \cite{cSurv} applied the permutation-based multiple-testing adjustment and built a web application providing a pipeline to segment cancer patients based on the optimal cutoffs of the gene expression biomarkers. However, the website suffers a low-speed issue due to the heavy computational load of permutation. It responds around $20$ seconds to a response, which is far from the good load time (e.g. $0-4$ seconds) recommended for the best website conversion rate \citep{portent}. In reality, since we must generate and analyze a lot of copies of permuted data to achieve a decent precision in the p-value, we cannot go around the computational issue. And this drawback led to its inefficiency in practical application. 

The drawbacks of existing methods motivate us to develop a novel method, which does not require investigating many candidate cutoffs to ensure the validity of asymptotic distribution and is much less computationally expensive than permutation approaches. Our novel approach borrow the idea from the literature on Group Sequential Design (GSD) of clinical trials. In the GSD, a series of interim analyses are conducted on datasets overlapped in a nested way, i.e. in any two interim analyses, the early dataset is a subset of the latter.  The GSD textbook \citep{group} (Page 2 of Chapter 11) introduced that the estimated coefficients from the same regression model fitted to the nested data follow the multivariate normal distribution, and their correlations can be explicitly calculated. The discussion covers various outcome types under their associated models,  including linear and generalized linear regressions in Chapter 3 and Cox regressions in Chapter 13. Moreover, Lausen and Schumacher \cite{Lausen1992} demonstrated that the log-rank statistics for survival outcome also follow the multivariate normal distribution under random (or non-informative) censoring. Note that these theoretical results in GSD are based on nested data. In the patient segmentation problem, the generated binary variables are overlapped but not in a nested way. This work will use theoretical derivations to extend GSD results to solve this patient segmentation problem, and propose a novel method for patient segmentation called BOSS (Biomarker Optimal Segmentation System). 

In BOSS, we model the joint distribution of the test statistics obtained from all cutoff values as a multivariate normal distribution, utilize the specific data overlapping structures of different cutoff values to calculate their correlation structures and precisely calculate the FWER from its definition without assuming any independence among the tests. Consequently, BOSS has two major advantages. First, it directly calculates the FWER from a closed-form formula, leading to a much faster speed than the permutation-based methods. Therefore, it is suitable to be incorporated into the web system, such as cSurvival \citep{cSurv}. Second, since it does not require many candidate cutoffs, BOSS can ensure decent statistical power, especially when the sample size is not big enough to investigate a lot of cutoffs.

The rest of this article is organized as follows. Section 2 sets up the problem and presents our model in detail. Section 3 conducts simulation studies to evaluate the performance of the BOSS method by comparing it with the permutation test. Section 4 illustrates our method via its application to real data. Section 5 includes conclusions and discussions.

\section{Methods}\label{sec2}
\subsection{Notation and Model}\label{proset}
Let vector $\mathbf{Y}=(Y_1, \ldots, Y_n)$ denote the clinical outcomes of $n$ patients and vector $\mathbf{B}=(B_1, \ldots, B_n)$ denote the measurements of a quantitative biomarker, where $n$ is the sample size. Let the $n$ by $(p+1)$ dimension matrix $\mathbf{C}=(\mathbf{1}, \mathbf{C}_1, \ldots, \mathbf{C}_p)$ denote the design matrix consisting of a column of $1$'s and other columns of $p$ covariate values (e.g. age, gender, and BMI, etc). We aim to segment patients into two groups based on a cutoff value for the biomarker. If the difference between these two groups' clinical outcomes is tested statistically significant, we claim the biomarker can be used to segment these patients. We investigate $k$ candidate cutoffs $(\tau_1, \ldots, \tau_k)$ to find the best one and test if it can segment patients into two groups with significantly different clinical outcomes after adjusting for the covariates. The $i$-th cutoff $\tau_i$ converts continuous biomarker $\mathbf{B}$ into a binary variable $\mathbf{X}_i=(X_{i1}, \ldots, X_{in})$ defined as 
\begin{align} \label{e:dummy}
    X_{is}= 
    \begin{cases} 
    1,& \mathrm{if }\; B_s>\tau_i\\
    0,& \mathrm{if }\; B_s \leq \tau_i.
    \end{cases} \;\; \mathrm{for }\; s=1,\ldots, n
\end{align}
For each $i$, we consider a linear regression model as below. 
\begin{align} \label{e:regression}
    \mbox{E}(\mathbf{Y}) = \beta_i \mathbf{X}_i + \gamma_{0i}+ \Sigma_{l=1}^p \gamma_{li} \mathbf{C}_l,
\end{align}
where $\beta_{i}$ represents the difference of clinical outcomes between two groups adjusted to other covariates, $\gamma_{0i}$ is the intercept, and $\gamma_{li}$ is the regression coefficient for the $l$th covariate. 

\subsection{Hypothesis Testing}
Based on the above model, we test the hypothesis of whether the biomarker can segment patients. The null hypothesis is the biomarker cannot segment patients at any given cutoff, i.e. 
\begin{equation} \label{e:nullh}
H_0: \;\; \beta_{1} = \beta_{2} = \cdots = \beta_{k} = 0. 
\end{equation}
The alternative hypothesis, $H_{1}$, is that at least one of these $\beta_{i}$s' is not equal to $0$. We consider the Z statistics, which follows standard normal distribution under $H_{0}$, i.e. 
\begin{align} \label{e:marginaldistn}
    Z_i  = \hat{\beta}_i/sd(\hat{\beta}_i) \sim N(0,1) \;\;\; \mathrm{under} \;\; H_{0} ,
\end{align}  
where $\hat{\beta}_i$ is the least squares estimate of the coefficient and $sd(\hat{\beta}_i)$ is its standard deviation. If we derive $k$ p-values from these standard normal distributions, they cannot be directly used to make a decision, since we must correct the inflated false positive probability from conducting multiple tests. 

Following the literature about maximumly selected statistics, we use the most extreme test statistics to make decisions, and show it is equivalent to FWER of conducting $k$ tests in our situation. We denote $z_i$ as the observed value of the Z-statistic, which can be obtained from the output of the fitted regression models, and will be used to calculate FWER. Under $H_0$, the joint distribution of $Z_i$'s is symmetric to zero in any dimension, so we only need to consider the absolute value of $z_i$'s under $H_0$. That is, when $z_i<0$, we have
\begin{align*}
    P(Z_i > z_i) = P(Z_i < |z_i|) \;\;\; \textrm{when $z_i<0$ and $H_{0i}$ is true.} 
\end{align*}
So, to simplify the derivation of FWER, we assume $z_i \geq 0$ for all $i=1, \ldots, k$; and we define the index of optimal cutoff $k^*$ as the one corresponding to the largest observed test statistics, 
\begin{align}\label{e:kbest}
k^* = \mathrm{argmax}\{z_i | i=1,\ldots,k\} 
\end{align}
the type I error of maximumly selected test statistics or the FWER of $k$ tests can be defined as below,
\begin{align}  
\nonumber FWER &= 1- P(\max_i Z_i \leq z_{k^*}| H_0)\\
\label{e:FWER}     &= 1- P(Z_1 \leq z_{k^*}, \ldots , Z_k \leq z_{k^*} | H_0).
\end{align}
We reject the null hypothesis $H_0$ if $\mathrm{FWER} < \alpha$, and fail to reject if $\mathrm{FWER} \geq \alpha$, where $\alpha$ is the significance level (usually $\alpha = 0.05$). To calculate the FWER in the Formula (\ref{e:FWER}), we need these Z-statistics $Z_i$'s joint distribution under the null hypothesis $H_0$, which follows a multivariate normal distribution\citep{group}
\begin{align}
 (Z_1, \ldots, Z_k) |H_0 \sim N(\mathbf{0}, \mathbf{\Sigma}).
\end{align}
This work's most challenging part is calculating the elements in the covariance matrix $\mathbf{\Sigma}$. We give its results below in formulas~(\ref{corr1}) and (\ref{corr}), and provide their detailed derivation in the supplementary document. 

The $(i,j)$ element of the covariance matrix $\mathbf{\Sigma}$ can be calculated as 
\begin{align} \label{corr1}
Cov(Z_i,Z_j)  =\frac{\mathbf{X}_{i}^T(\mathbf{I}-\mathbf{H})\mathbf{X}_{j}}{\sqrt{\mathbf{X}_{i}^T(\mathbf{I}-\mathbf{H})\mathbf{X}_{i}}\sqrt{ \mathbf{X}_{j}^T(\mathbf{I}-\mathbf{H})\mathbf{X}_{j}}}
\end{align}
where $\mathbf{H}=\mathbf{C}(\mathbf{C}^T\mathbf{C})^{-1}\mathbf{C}^T$ is hat matrix of covariates. If no covariates exist in the regression model, the covariance in Formula~(\ref{corr1}) is reduced to a simpler form
\begin{align}\label{corr}
 Cov(Z_i,Z_j)= \sqrt{(n-m_{j})m_{i}}/\sqrt{(n-m_{i})m_{j}},  
\end{align}
where $m_{i} \leq m_{j}$, $m_{i}$ and $m_{j}$ are the numbers of $1$s' in vectors $\mathbf{X}_{i}$ and $\mathbf{X}_{j}$ respectively, and $n$ is the total number of samples.  Formula~(\ref{corr}) agrees with our intuition that without covariates in the model, the correlation between the estimated regression coefficients from models associated with two different cutoffs only depends on the overlapping $1$'s proportion in these two derived binary variables.

\begin{algorithm}
\caption{Select optimal cutoff and test significance} 
\label{alg:Framwork} 
\begin{algorithmic} 
 \Require
 Clinical Outcome $\mathbf{Y}$; Covariates $\mathbf{C}$;  A Continuous Biomarker $\mathbf{B}$; Candidate Cutoffs $(\tau_1,\dots, \tau_k)$; Significance Level $\alpha=0.05$
\renewcommand{\algorithmicrequire}{ \textbf{Procedure:}}
\Require 
\State 1. \textbf{For}  $i$ in $1, \ldots, k$  \textbf{do}  
\State (a) Derive $\mathbf{X}_i$ based on the $i$th cutoff $\tau_i$ and biomarker measurements $\mathbf{B}$ using Formula~(\ref{e:dummy}).
\State(b) Regress $\mathbf{X}_i$ on $\mathbf{Y}$ with adjustment to $\mathbf{C}$ in   a regression model to obtain the estimated regression coefficient $\hat{\beta}_i$, its estimated standard deviation $\hat{sd}(\hat{\beta}_i)$, and the Z-statistics $z_i$.\\
\textbf{end for}
\State 2. Find the index of the optimal cutoff, $k^*=\max_k(|z_1|, \ldots, |z_k|)$.
\State 3. Calculate the components of $Z_i$'s covariance $\mathbf{\Sigma}$, using Formulas~(\ref{corr1}) or (\ref{corr}). 
\State 4. Calculate FWER using Formula~(\ref{e:FWER}).
\State 5. If FWER $<\alpha$, reject $H_0$ and claim that using biomarker and its cutoff $\tau_{k^*}$ can segment patients into significantly different groups.
\Ensure
optimal cutoff $\tau_{k^*}$ and its FWER value\\
\end{algorithmic} 
\end{algorithm}

Finally, we summarize the analysis procedure of BOSS in Algorithm~\ref{alg:Framwork}. 


\subsection{Extension to other types of regression models}
In the GSD textbook \citep{group}, the correlation formula of Z-statistics among the nested interim analyses is the same for linear, generalized, and Cox regression, which shines the light on the belief that the same property should hold for other types of regression. In this work, we skip the mathematical derivation for other regression models and use simulation studies to demonstrate that formula~(\ref{corr1}) is also valid in Cox regressions.

\section{Simulation Studies}
The main purpose of our simulation studies is to compare the performance of our novel method, BOSS, with the gold standard permutation method. We simulate data under various settings and apply both methods to the data to compare their statistical powers, Type I errors, and computation times. The power measures the ability of the tool to identify true signals, while the type I error indicates false discoveries. An optimal tool is expected to have high power and properly control type I error to a nominated threshold ($0.05$ being the commonly used value ). The computation time is essential in many applications, especially for website-based tools and when we must repeat the analysis many times (e.g. for tens of thousands of genes). 

\subsection{Simulation Procedure} \label{simulation}
To ensure the simulated data closely mimic the real-world data, we employ real genomic data to build a blueprint model and simulate outcome variables based on such blueprint models while reserving genomic information and patients' other characteristics in the real data. In particular, we use data from a curated cancer outcome database \citep{cSurv}, containing expression levels of $17,431$ genes of $500$ lung adenocarcinoma cases. The detailed background of the data is illustrated in the next section of the data application. We carry out simulations in the following two steps.

\noindent\textbf{Step 1: Build Blueprint Model}
We apply BOSS to the $17,431$ genes separately in the data mentioned above and obtain their FWERs defined in the Formula (\ref{e:FWER}), which results in $2,089$ identified significant genes via linear regression model fitting and $3,502$ identified significant genes via Cox regression model fitting based on the significance threshold FWER$<0.05$. From the identified genes, we select the top $50$ genes with the smallest FWERs (i.e. the top 50 strong-effect biomarkers associated with the outcome) and $50$ genes with the largest FWERs (i.e. the bottom 50 weak-effect biomarkers associated with the outcome). In total, we obtain $100$ representative genes for linear regression and Cox regression, respectively. To describe the blueprint model, we denote the expression levels of these selected genes as $\mathbf{B}_g$ and $\tau^*_g$ as the $g$-th gene's optimal cutoff selected by BOSS with $g=1, \ldots, 100$. Let $\mathbf{Y}_g$ be the vector of the outcome variable (quantitative or survival types). For quantitative outcomes,
\begin{equation} \label{e:simlin1}
E(\mathbf{Y}_g) = \beta_{_g} I\{\mathbf{B}_g > \tau^*_g\} +\gamma_{0g},
\end{equation}
and for survival outcomes,
\begin{equation} \label{e:simlin2}
h(\mathbf{Y}_g ) = h_{0}(\mathbf{Y}_g)\mbox{exp}\left(\beta_{_g} I\{\mathbf{B}_g > \tau^*_g\} +\gamma_{0g}\right),
\end{equation}
where $\beta_{_g}$ is the association parameter between the $g$th gene (biomarker) and the outcome variable, $\tau^*_g$ is the optimal threshold, and $\gamma_{0g}$ is the corresponding linear component associated with the covariates ($g=1,...,100$). 

In summary, based on $100$ selected genes, we obtained $100$ blueprint models for the linear regression models and $100$ for Cox regression models.

\noindent\textbf{Step 2: Simulate Data}
All simulated data retain genomic information from original real data, and simulate clinical outcomes using the blueprint models. We simulate quantitative outcomes using the Gaussian random sampler and the model (\ref{e:simlin1}), and simulate survival outcomes using the R package \textit{coxed} \citep{simsurv} and the model (\ref{e:simlin2}). This R package uses a random spline method to generate a baseline hazard function via a cubic spline fitting at randomly drawn points. 

For each blueprint model discussed above, we simulate $100$ copies of positive data and $100$ copies of negative data. We denote the positive data as $(\mathbf{Y}_g, \mathbf{B}_g)$. The outcomes of positive data are generated using parameters of blueprint model learned from the original data, whose true association parameters $\beta_{_g}$s' are non-zero values. Hence, the proportion of identified significant biomarkers in data analysis represents the statistical power. We denote the negative data as $(\mathbf{Y}^0_g, \mathbf{B}_g)$. The outcomes of negative data are generated using zeroed-out association parameters (i.e. setting $\beta_{g}=0$) in blue print models. Hence, the generated outcomes are unrelated to biomarkers, and the proportion of identified significant biomarkers in data analysis represents the type I error.

\subsection{Analysis Scheme}
We apply BOSS to analyze each copy of $(\mathbf{Y}_g, \mathbf{B}_g)$ and $(\mathbf{Y}^0_g, \mathbf{B}_g)$ separately. We claim the $g$-th gene is significant, if its BOSS FWER (\ref{e:FWER}) is less than $0.05$. Note that the analyses are repeated with various values of the number of candidate cutoffs, $k= 6, 8, 10, 12, 14$, and we choose the cutoff points $(\tau_1, \ldots, \tau_k)$ as equally spaced quantiles of observed biomarker values. In this way, we obtain $100$ FWERs for every gene from each simulated data under various $k$ and two outcome types. 

As for evaluation metrics, given a $g \in \{1, 2, \cdots, 100\}$, a $k \in \{6, 8, 10, 12, 14\}$ and one outcome type (i.e. quantitative or survival outcome), we use the proportion of FWER$<0.05$ from these $100$ copies $(\mathbf{Y}_g, \mathbf{B}_g)$ to represent the statistical power of the BOSS test and use the proportion of FWER$<0.05$ from these $100$ copies $(\mathbf{Y}^0_g, \mathbf{B}_g)$ to represent the Type I error rate of the BOSS test under such condition.    

We repeat the above analyses using a permutation approach with $1000$ permutations to achieve the precision of $0.001$ in each estimated FWER, which results in the corresponding pairs of estimated statistical power and Type I error, based on the same rules for BOSS.

\subsection{Simulation Results}

\subsubsection{Comparisons based on Evaluation Metrics}
\begin{figure}[htbp]
\centering
\includegraphics[width=8cm,height=7cm]{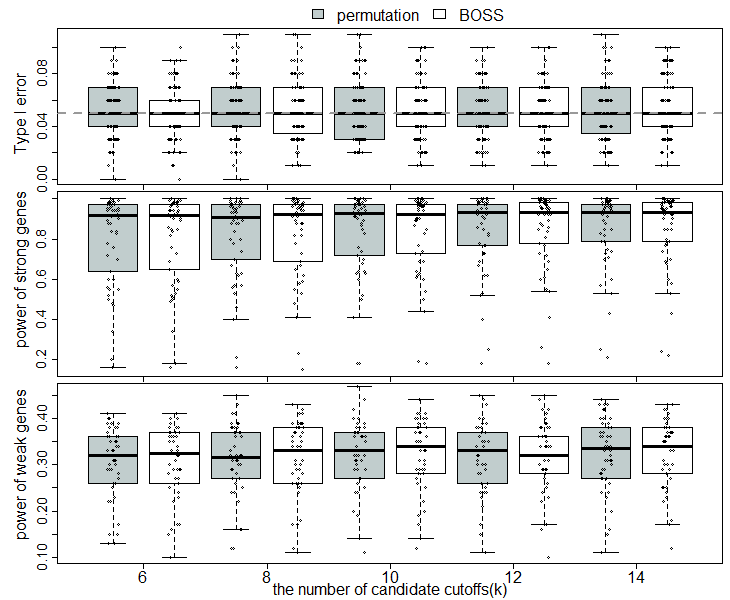}
\caption{Boxplots of type I error rate and power from simulated data with quantitative outcomes. In the top panel, the boxplots of type I error rates of both methods under various $k$, the number of candidate cutoffs, are presented. In the middle and bottom panels, the boxplots of the power of both methods based on the $50$ strong biomarkers and $50$ weak biomarkers under various $k$ are presented separately.}  \label{f:linear}
\end{figure}
\begin{figure}[htbp]
\centering
\includegraphics[width=8cm,height=7cm]{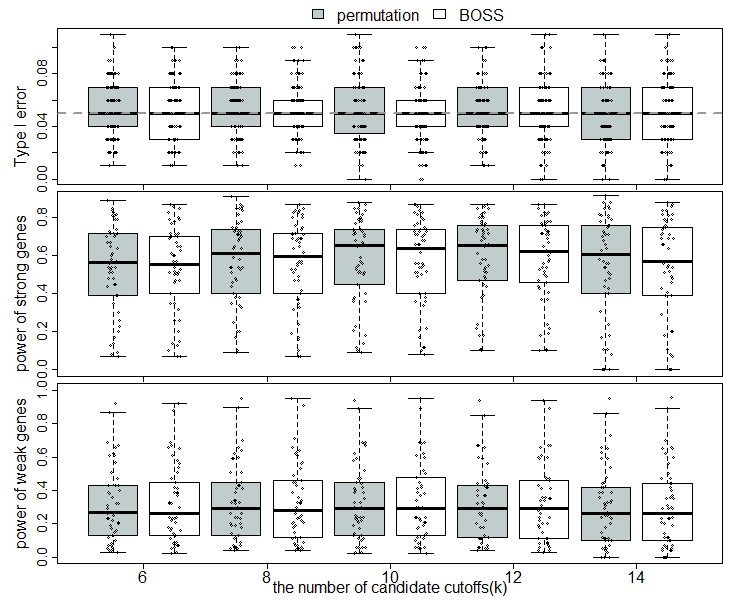}
\caption{Boxplots of type I error rate and power from simulated data with survival outcomes. In the top panel, the boxplots of type I error rates of both methods under various $k$, the number of candidate cutoffs, are presented. In the middle and bottom panels, the boxplots of the power of both methods based on the $50$ strong biomarkers and $50$ weak biomarkers under various $k$ are presented separately.}
\label{f:surv}
\end{figure}
We employ side-by-side boxplots of the resulting power and type I error to demonstrate the performance of the BOSS and permutation methods in Figure~\ref{f:linear} for quantitative outcomes and Figure~\ref{f:surv} for survival outcomes separately. 

First, in the top panel of each figure, boxplots of type I errors of the two methods are presented under various $k$, the number of candidate cutoffs, with a dashed horizon line indicating the nominated threshold $0.05$ for controlling type I error rate. These boxplots indicate the type I errors of both BOSS and the permutation tests are properly controlled around $0.05$ (i.e. centred at $0.05$ with most of IQRs being less than $0.04$). The one-sample Wilcoxon tests are conducted to numbers in each box to compare these resulting type I errors against the nominated level of $0.05$. We found none of these boxes is significantly different from $0.05$ after the Bonferroni correction. Therefore, both methods perform well in controlling false positives at the desired level. 

Second, the middle panels show the medians of the powers obtained from the $50$ strong-effect genes of both BOSS and the permutation method are all above $80$\% for quantitative outcomes (Figure \ref{f:linear}) and all close to $60$\% for survival outcomes (Figure \ref{f:surv}) under various simulation settings, which indicates both methods have similar decent powers under all the simulation settings for biomarkers with strong effects. When the biomarkers carry weak signals, the medians of the powers from both methods decrease to around $30$\% for quantitative outcomes (Figure \ref{f:linear}) and $20$\% for survival outcomes (Figure \ref{f:surv}) separately. 

Based on the visual comparison from boxes of BOSS versus permutation approach, we found these two methods perform very similarly. We confirmed the visual pattern using the paired Wilcoxon tests, and found no significant p-value after the Bonferroni correction. Therefore, we conclude BOSS performs equally well as the permutation approach regarding statistical power and Type I error control.      

\subsubsection{Comparisons based on Computation Time}
\begin{figure}[htbp]
\centering
\includegraphics[scale=0.26]{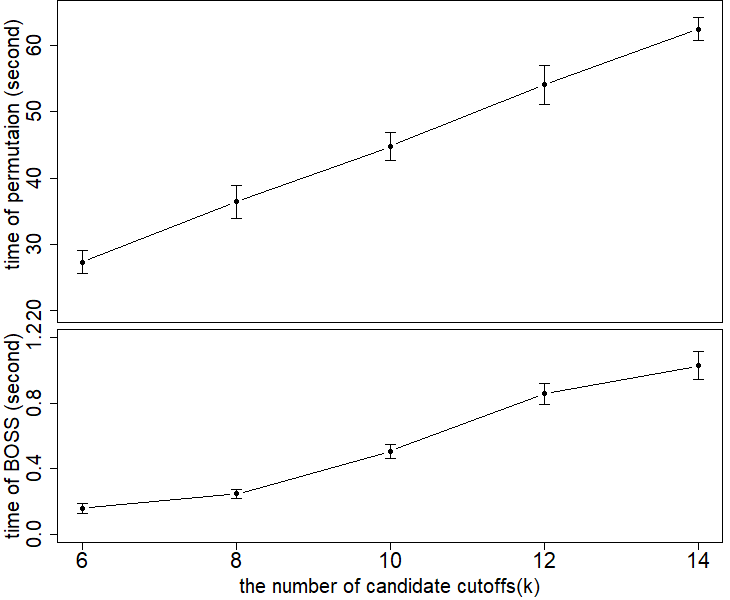} 
\caption{ Summary of computation time of the BOSS and the permutation methods in simulation studies. The mean and the corresponding one-standard-error bar of computation time for each method are aggregated based on the pooled computation time under each $k$, the number of candidate cutoffs, from the simulation studies.}
\label{time}
\end{figure}
We summarize the computation time of both methods from simulation studies and present them in Figure~\ref{time}. It clearly shows that BOSS is much faster than a permutation approach. In particular, BOSS decreases the computation time by $175$, $146$, $90$, $64$, $60$ times when $k = 6, 8, 10, 12,$ and $14$ correspondingly. Furthermore, although both methods' computation time increases with the number of candidate cutoffs, the ratio of the two methods is gradually levelled off.

Note that the permutation approach carried out in the simulation study only consists of $1000$ permutations, providing the estimate of the FWER with the precision of $0.001$ (i.e. three digits from the decimal point). The precision of these estimates is often not enough in many applications. If we need to increase the precision of the FWER by one more digit from the decimal point, the number of required permutations increases to $10$ folds, leading to at least a $10$ fold increase in the computing time of the permutation approach. In contrast, BOSS estimates FWER theoretically and its precision does not interfere with the computing time.

\subsection{Sensitivity Analysis on Sample Size}
As a sensitivity analysis, we explore if the sample size has an effect on the performance of the two methods. The previous simulation studies are based on the same sample size with the real data employed to generate blueprint models. So in this analysis we generate data with different sample sizes. In particular, we conduct random sampling with replacement to these simulated datasets separately to generate one with half size and the other with double size from each original dataset. After that we apply the analysis scheme to these generated data and compare their results with the ones from the original data to make conclusions. 

\begin{figure}[htbp]
\centering
\includegraphics[width=7.5cm,height=5.5cm]{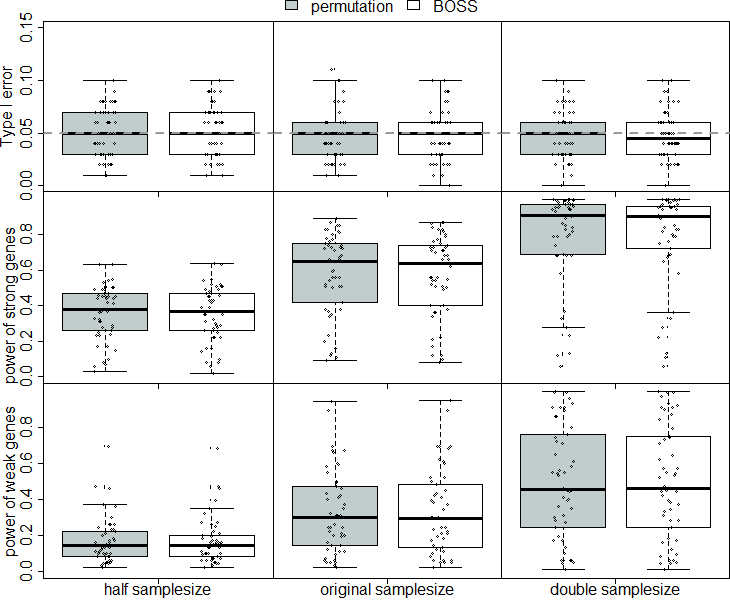} %
\caption{ Boxplots of type I error rate and power from sensitivity analysis. Based on survival outcome under $k=10$, boxplots from the half, original, and doubled sized data are presented in the left, middle and right penals seperately. And the type I error rates and two types of power from the top $50$ and the bottom $50$ biomarkers are presented in the top, middle and bottom panels separately. Boxplots from BOSS are filled with white colors, while boxplots from the permutation method are filled with grey colors for comparison.}
\label{f:size}
\end{figure}
The sensitivity analysis results confirm that under different sample sizes, BOSS and permutation perform very similarly in statistical power and Type I error. For example, Figure~\ref{f:size} shows the distributions of estimated power and type I error rates from further investigations under a fixed $k=10$ and survival outcomes with various-sized data. Furthermore, these results indicate that increasing the sample size can largely improve the statistical power for both BOSS and permutation. However, changes in sample sizes do not affect the type I error rates.

In summary, simulation studies demonstrate that the novel BOSS method works as good as the gold standard permutation method regarding the ability in detecting positive signals and controlling false positive discoveries, which are not affected by sample size. However, BOSS uses much less time than the permutation method, satisifying the critieria of a good load time for a webtool and is more useful when we need to repeat analysis many times (e.g. analyze many genes in a genomic study).

\section{Real Data Application}
Lung cancer is the leading cause of cancer death worldwide \citep{Siegel}. Lung adenocarcinoma (LUAD) is a type of lung cancer that develops in the epithelial cells of the lungs. Drug and immune resistance has been extensively reported in LUAD \cite{Spella,Iglesias}. Revealing the genetic mechanisms underlying LUAD can facilitate understanding LUAD etiology and hence offers new entry points for treatment and/or disease management. To demonstrate the novel method provides new insights into the cancer prognosis, we employed the BOSS to analyze the curated LUAD genomic and clinical data published in \cite{cSurv}, which contains the expressions of $17431$ genes collected from $500$ LUAD cases and their corresponding clinical information including gender, overall survival time, and censoring status. This dataset can be downloaded from the website provided in the Data availability statement.


The results from BOSS provide the optimal cutoff and the FWER (i.e. the adjusted p-value defined in Formula \ref{e:FWER}) for each gene. Note that the FWER makes a correction in BOSS to avoid p-value inflation due to multiple testing of different cutoff values for the same gene. In this analysis, we conducted analysis for $17431$ genes, which require another multiple testing adjustment. We further corrected these $17431$ BOSS FWERs to control the False Discovery Rate (FDR) of testing many genes at level $0.05$. After the FDR adjustment, we found expressions of $48$ genes having optimal cutoffs and showing significant (FDR-adjusted $p < 0.05$) correlation with overall survival time in LUAD (Supplementary Table S1). We performed a literature review of these $48$ genes and found that $20$ of them were specifically reported to be associated with LUAD (Supplementary Table S1). For example, \cite{Xu} demonstrated that overexpression of ANLN is associated with metastasis of LUAD complications. Liu et al. \cite{Liu} showed that CXCL induced spinal metastasis of LUAD. Xu et al. \cite{Wu} identified FAM117A as a novel prognostic marker for poor outcomes in lung cancer patients. The supports from independent evidence from the literature demonstrate that BOSS can find useful prognostic genes and generate biological insights. There are $28$ genes discovered by our method but have no literature support. We consider them as potential novel genes and suggest researchers further investigate and validate them.

\begin{figure}[htbp]
\centering
\includegraphics[scale=0.36]{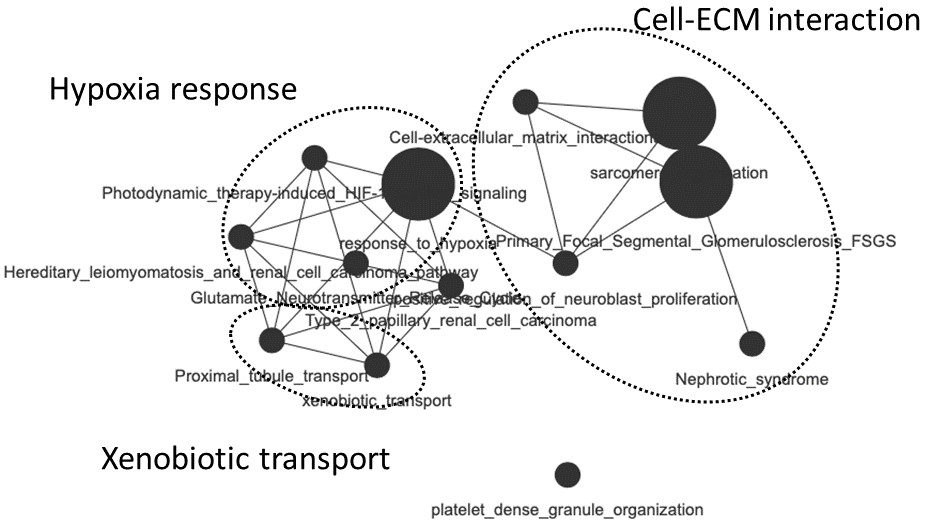} %
\caption{Enrichment network of overrepresented (pval $\textless$ 0.0001, padj $<$ 0.075) GSs in genes correlated with overall survival in lung adenocarcinoma. Node denotes GS; node size reflects the number of genes in each GS. Edge reflects significant gene overlap as defined by a Jaccard Coefficient larger than or equal to 0.25. Detailed statistics are provided in Supplementary Table S2. ECM, extracellular matrix; GS, gene set; padj, adjusted P-value; pval, P-value.}\label{f:enrich}
\end{figure}

To understand the common messages in these $48$ genes, we performed functional enrichment analysis with eVITTA \citep{eVITTA} (Supplementary Table S2). We found that these $48$ genes are overrepresented in cancer hallmark pathways \citep{Han}, including cell-extracellular matrix (ECM) interaction, xenobiotic transport, and hypoxia response (Figure~\ref{f:enrich}). Notable genes include ITGB1, FLNC, SLC2A1, and EGLN1. Among them, ITGB1, $\beta$1 integrin, is the predominantly expressed integrin protein in normal and tumor cells playing a pivotal role in ECM-related signaling; its expression is found to be prognostic in non-small-cell lung cancer \citep{Weiqi}. FLNC, or filamin-C, as an important cytoskeleton component, plays a key role in cell migration and signal transduction \citep{Ai}. While its role in lung cancer is not extensively reported, it's been shown to be prognostic in other cancers such as prostate \citep{Ai}. SLC2A1 or glucose transporter 1, and EGLN1 or hypoxia-inducible factor prolyl hydroxylase 2, are two key mediators of hypoxia responses \citep{Dengler}. Hypoxia is an important feature lung cancer microenvironment related to cancer progression, metastasis, and metabolism \citep{Iwona}. In hypoxia responses, SLC2A1 helps transition from oxidative towards glycolytic metabolism in cancer cells while EGLN1 modifies HIFs Post-translationally \citep{Dengler}. This network highlights potential converging mechanisms that might influence malignancy synergistically, advancing LUAD and promoting resistance to treatment.

Together, our method was robust in detecting potentially prognostic biomarkers by examining all expressed genes with reasonable computational resources and time. The biomarkers identified with BOSS in this work have referential importance in exploring synergy and target-matched therapies in LUAD.

\section{Conclusion and Discussion}
Research on selecting the optimal cutoffs is conducted in various fields, such as machine learning, biology, and medicine. In machine learning, \cite{Shih, Lausen} used the optimal cutoff selection in building decision and regression trees. In biology, optimal cutoffs are utilized to distinguish species abundance based on ecological factors \citep{Miiller}.  In clinical trials, the optimal cutoffs are used in different settings, such as assessing the association between the characteristics of tumor cells and the prognosis of patients \citep{Galon}, and obtaining the threshold for residual leukemic cells in acute myeloid leukemia patients \citep{Buccisano}. However, these proposed methods cannot be not directly applied to our case, i.e. featuring a special overlapping data structure generated from several candidate cutoffs. Therefore, we developed our novel statistical test method, BOSS, which has three main advantages: fast calculating speed, high accuracy, and no requirement for a large number of candidate cutoffs. First, BOSS has decent speed, leading to great practical significance. \cite{weng} shows that every 0.4-second increase in Google page load time reduces at least 8 million daily searches. And every 1-second rise in Amazon's Web load time reduces at least $1.6$ billion company's annual revenue. The speed of BOSS satisfies the good load time criteria. Therefore, it can greatly improve the user experience if it replaces the current permutation method used in the web system for medical analysis \citep{cSurv}. Second, BOSS is accurate. And its accuracy is consistent and does not interfere with the speed. In contrast, the accuracy of the permutation test is proportional to the number of permutations and inversely proportional to the speed. High accuracy requirement in permutation leads to severly prolonged running time. Third, BOSS does not require a large number of candidate cutoffs. BOSS is designed based on controlling family wise error rate, not requiring a large number of hypothesis testings. Furthermore, we carry out theoretical derivation to calculate the adjusted p values from the exact false positive probabilities based on specific data overlapping structures, which improve the accuracy and remove the requirement of large sample asymptotic properties.       

In practice, the number of candidate cutoffs is related to the power in two contradictory directions. On one side, investigating more cutoffs enables finding an optimal value closer to the truth, which increases the power. On the other hand, investigating more cutoffs requires more tests causing heavier penalties in the multiple testing adjustment, which decreases the power. Our simulation studies provide some evidence and insights into that paradox. For quantitative outcomes in Figure~\ref{f:linear}, most powers of genes with strong signals are above $80\%$, and the best median power is achieved at the maximum number of cutoffs $k=14$, whereas most power of genes with weak signals are between $30\%$ and $40\%$ and the best median power is achieved at $k=10$. For survival outcomes, most powers are $60\%$ or lower, and the best median powers are not achieved at the max value of $k$. This suggests exploring more candidate cutoffs to find a more precise optimal cutoff that can better segment patients, when the signal is strong enough to achieve a good power of $80\%$; and suggests reducing the number of cutoffs to lower the penalty of multiple testing adjustments, when the signal is not strong to achieve a good power.  In summary, when selecting the number of candidate cutoffs, we should balance between two factors to achieve the best results: (1) find a more precise optimal cutoff and (2) reduce multiple testing penalties.

In this article, we discuss the method to find the optimal cutoffs with one biomarker, which is the most common case in practice, i.e., only using one to segment patients into subpopulations. However, sometimes researchers need to consider more than one biomarker. For example, researchers who are interested in two biomarkers often explore differences between samples with double positive or double negative biomarkers (i.e., subgroups where biomarkers are both high or low). Patients with two biomarkers’ values in opposite directions (one protective and one risk) are excluded from personalized medicine in this situation due to the difficulty of defining their group membership. We can convert two continuous biomarkers into a binary variable $X_i$ and apply BOSS to compare double positives versus double negatives. The detailed process of transforming the two-biomarker problem into a one-biomarker problem is discussed in the Supplementary document with a numerical example. 

In summary, BOSS is a powerful tool for selecting optimal cutoffs in various research fields, especially in the scenario requiring fast computing speed to develop a web application or needing to repeat analysis many times (e.g. on a vast number of biomarkers in genomic studies). The derivation of BOSS inspires researchers to develop multiple test adjustments based on specific data overlapping structures.

\section*{Supplementary document}
 The supplementary document includes 4 parts: (1) the derivation of the covariance formula of Z-statistics in linear regression models, (2)scenarios of finding optimal cutoffs for multiple variables,  (3) the table of 48 genes with significant optimal cutoffs, and (4) overrepresented genes in cancer hallmark pathways. 
 

\section*{Data availability statement}\label{data source}
The cancer patient survival data were obtained from the cSurvival program \citep{cSurv}. Raw survival data can be downloaded from https://tau.cmmt.ubc.ca/cSurvival/project\_data/TCGA-LUAD/df\_survival\_o.csv. Raw gene data can be downloaded from https://tau.cmmt.ubc.ca/cSurvival/project\_data/TCGA-LUAD/df\_gene.csv. 

\bibliographystyle{apa}
\bibliography{boss}


\end{document}